\documentclass[12pt,preprint]{aastex}
\shorttitle{Planets Around Massive White Dwarfs}
\shortauthors{Kilic, Gould, \& Koester}

\begin{document}

\title{Limits on Unresolved Planetary Companions to White Dwarf Remnants of 14 Intermediate-Mass Stars}

\author{Mukremin Kilic\altaffilmark{1,4}, Andrew Gould\altaffilmark{2}, and Detlev Koester\altaffilmark{3}}

\altaffiltext{1}{Smithsonian Astrophysical Observatory, 60 Garden Street, Cambridge,
MA 02138, USA}
\altaffiltext{2}{Department of Astronomy, Ohio State University,
140 W.\ 18th Ave., Columbus, OH 43210, USA}
\altaffiltext{3}{Institut f\"ur Theoretische Physik und Astrophysik, University of Kiel, 24098 Kiel, Germany}
\altaffiltext{4}{Spitzer Fellow; mkilic@cfa.harvard.edu}

\begin{abstract}

We present {\em Spitzer} IRAC photometry of white dwarf remnants of 14 stars with $M= 3-5\ M_\odot$.
We do not detect mid-infrared excess around any of our targets. By demanding a 3$\sigma$ photometric excess
at 4.5 $\mu$m for unresolved companions, we rule out planetary mass companions down to 5, 7, or 10 $M_J$
for 13 of our targets based on the \citet{burrows03} substellar cooling models.
Combined with previous IRAC observations of white dwarf remnants of intermediate-mass stars,
we rule out $\geq10M_J$ companions around 40 white dwarfs and $\geq5M_J$ companions around
10 white dwarfs.

\end{abstract}


\section{INTRODUCTION}

Radial velocity, transit, and microlensing searches are succesful in finding planets
around stars less massive than $2 M_\odot$ \citep[see][and references therein]{udry07,marcy05,gould09,mazeh09}.
However, these techniques have limitations for higher mass stars. The main problem is that intermediate- and
high-mass stars are big, they are rare, and they do not have many usable lines for radial velocity
studies. The short ($\leq1$ Gyr) main-sequence lifetimes of intermediate-mass stars 
and the small radii of the remnant white dwarfs (WDs) imply that the planetary systems
around massive stars can be effectively studied after the hosts have been transformed into WDs
\citep{ignace01,burleigh02}.

Compared to main-sequence stars, the significant gain in contrast makes WDs excellent targets
for photometric searches for planetary companions. Massive WDs are smaller and fainter than average mass ($0.6M_\odot$)
WDs, increasing the contrast for substellar companions even further.
If planetary systems survive the late stages of stellar evolution, massive planets should be detectable around
WDs \citep{burleigh02}.

Detailed dynamical simulations of the evolution of the Solar system by \citet{duncan98} show that the giant planets
are likely to remain in stable orbits for more than 10 Gyr after the Sun becomes a WD. 
The discovery of an $M\sin i = 3.2 M_J$ planet around the sdB star V391 Pegasi by \citet{silvotti07} and a candidate
$\geq2.4 M_J$ planet around the WD GD 66 by \citet{mullally09} indicate that at least some planets survive
post main-sequence evolution.

There are now more than a dozen WDs known to host circumstellar debris disks \citep[see][and references therein]{farihi09,jura09,kilic08}.
These disks are
most likely formed by tidally disrupted asteroids \citep{jura03}, and at least in one case (GD 362),
the composition of the
accreted matter most resembles that of the Earth/Moon system \citep{zuckerman07}.
The atmospheric composition of other
disk-polluted WDs has not yet
been analyzed in such detail.
\citet{farihi09} estimate that at least 1\% to 3\% of WDs with cooling ages less than about 0.5 Gyr harbor debris disks.
There is indirect evidence from these debris disks that at least a few percent of WDs have remnant planetary systems
\citep{jura03,jura06}.

\citet{laws03} and \citet{fischer05} find a significant trend in the frequency of planets with increasing stellar mass.
The frequency of Jovian-planets ($M>0.8 M_J$ and $a<2.5$ AU) increases from 4\% for Sun-like stars to 9\% for $1.3-1.9
M_\odot$ stars \citep{johnson07}. Of course, the analogs of all of these planets around higher mass stars would most
likely be
swallowed in the asymptotic giant branch (AGB) phase \citep{villaver07}, and the frequency of planets around massive stars at wider
separations is poorly constrained at present. In the planet formation models of
\citet{kennedy08}, the fraction of stars with giant planets shows a steady increase with mass up to $3M_\odot$.
In addition, the mass of the planets and the width of the regions where they form are predicted to increase with stellar mass.
Observations of massive WDs can test these models. 

\citet{gould08} identified 49 young ($\leq1$ Gyr) WDs from the Palomar-Green Survey that are suitable for a mid-infrared search
for planetary mass companions. Here we present {\em Spitzer} IRAC observations of 14 stars
from that sample. Our observations are discussed in Section 2, while the spectral energy distributions and the limits
on planetary companions are discussed in Section 3.

\section{OBSERVATIONS}

We selected our targets from the DA WDs found in the Palomar-Green Survey \citep{liebert05}. The advantage
of this sample selection is that \citet{liebert05} provide $T_{\rm eff}, \log g$, mass, cooling
age, and distance
estimates for all these stars. We selected the 14 brightest WDs with $M_{WD}=0.7-0.9 M_\odot$ from that sample. 
The physical parameters of our targets, including the initial-mass estimates derived using the initial-final mass
relation of \citet{kalirai08}, are presented in Table 1. The initial-masses of our targets range from 2.8 to
4.6 $M_\odot$. The use of different initial-final mass relations results in a $\approx$10\% difference in the initial-mass
estimates.
We estimate the main-sequence (MS) lifetimes using the equation $t_{\rm MS} = 10 (\frac{M_{MS}}{M_\odot})^{-2.5}$ Gyr
\citep{wood92}. The total (WD + MS) ages of our targets are also presented in Table 1. Four targets are about
300 Myr old, and the remaining targets have ages $\approx1$ Gyr.

Observations reported here were obtained as part of M. Kilic's Cycle 5 {\em Spitzer}
Fellowship Program 474.
We obtained 3.6, 4.5, 5.8, and 7.9
$\mu$m images with integration times of 30 or 100 seconds per dither,
with five or nine dithers per object. We use the IRAF PHOT and IDL astrolib packages to perform
aperture photometry on the individual BCD frames from the latest available IRAC pipeline reduction.
Since our targets are relatively faint, we use the smallest aperture (two pixels) for which there are published
aperture corrections. Following the IRAC calibration procedure, corrections for the location of the source in the array
are taken into account before averaging the fluxes of each of the dithered frames at each wavelength.
Channel 1 (3.6$\mu$m) photometry is also corrected for the pixel-phase-dependence.
The results from IRAF and IDL reductions are consistent within the errors.
The photometric error bars are estimated from the observed scatter in the five (or nine) images
corresponding to the dither positions. We also add the 3\% absolute calibration error in quadrature.
Finally, we divide the estimated fluxes by the color corrections for a Rayleigh-Jeans spectrum \citep{reach05}.
These corrections are 1.0111, 1.0121, 1.0155, and 1.0337 for the 3.6, 4.5, 5.8, and 7.9
$\mu$m bands, respectively.

We present the IRAC photometry of our targets in Table 2. All of our targets are detected in 2MASS, at least in the
$J$ and $H$ bands. All but two of our targets also have Sloan Digital Sky Survey (SDSS) photometry available. 
PG 0852+659 and PG 1335+701 are not covered in the SDSS Data Release 7 area. We obtained $V$ and $I$ band photometry
of PG 1335+701 using the MDM 2.4m telescope equipped with the Echelle CCD. Two sets of 120 s exposures were obtained
in each filter on UT 2009 April 5. We use observations of the standard star field PG 1323$-$086 \citep{landolt92}
to calibrate the photometry. The photometric reductions were performed by J. Thorstensen, and kindly made available to us.
PG 1335+701 has $V=$ 15.29 mag and $I=$ 15.54 mag. We estimate an internal accuracy of 0.01 mag for the photometry,
but to be conservative, we adopt errors of 0.03 mag.
\citet{liebert05} provide an estimate on the $V$ magnitude of PG 0852+659 using photographic $B-$band photometry. However,
the error in the photographic magnitude is larger than 0.3 mag, and therefore we do not use it in our analysis.
PG 0852+659 is the only WD in our sample without accurate optical photometry.

\section{RESULTS}

Figure 1 presents the spectral energy distributions of four WDs with ages $\approx$ 300 Myr.
This youth is important; young planets will be brighter than their older counterparts
and therefore it should be easier to find planets around relatively young WDs.
The SDSS photometric zero points differ slightly from the AB convention \citep{eisenstein06}. 
We use the corrections given in \citet{eisenstein06} to convert the photometry to the AB system.
We calculate synthetic spectra for our targets using the model atmosphere code described by \citet{koester09} and
the best-fit $T_{\rm eff}$ and $\log g$ values from \citet{liebert05}.
We perform synthetic photometry on these WD models using the appropriate transmission
curves for the SDSS, 2MASS, and IRAC filters. The resulting fluxes are then compared with the observations to find
the normalization factor for the WD models. We weight fluxes by their associated error bars.
The solid lines in Figure 1 present the appropriate WD model for each star, which is
normalized to match the observations.
Optical and near-infrared photometry helps us to constrain the predicted mid-infrared photospheric fluxes
for WDs. The 4.5 $\mu$m photometry of all four WDs presented in this figure is consistent with the predicted
photospheric flux from WDs within 1$\sigma$; none of them show excess mid-infrared flux.

We use the synthetic spectra for 300 Myr old 1-25 $M_J$ planets \citep{burrows03} to put upper limits on
companions that would escape detection.
The overall agreement between the \citet{hubeny07} models and the spectral energy distributions of cold
brown dwarfs in the T dwarf range suggests that the planet models that we use are appropriate for
the colder planets expected around WDs.
The models by A. Burrows agree reasonably well with the observed
hot Jupiter emission spectra in the {\em Spitzer}
IRAC bands \citep[][H. Knutson 2009, priv. comm.]{burrows08}.
However, the predicted fluxes can vary by a factor of two due to temperature inversions in the atmosphere
and water being observed in emission or absorption.
These temperature inversions would make the infrared fluxes higher,
and they would
make it easier to find planets.

The dotted lines in Figure 1 show the combined flux from each WD plus 10, 7, 5, 2, and 1 $M_J$
companions \citep[from top to bottom,][]{burrows03}, respectively. The red cross marks the 3$\sigma$
upper limit of the 4.5 $\mu$m photometry.
By demanding a 3$\sigma$ excess from any possible companion, we exclude $\geq 5 M_J$ planets around PG 1038+634,
PG 1051+274, and PG 1446+286. We exclude $\geq 10 M_J$ companions around PG 1335+701. Of course, these mass limits
are model-dependent.

Figure 2 presents the spectral energy distributions of the remaining 10 WDs in our sample. All of these WDs are older than
about 600 Myr. We use 1 Gyr old planet models to constrain the contribution from possible companions.
As in Figure 1, the solid lines show the best-fit WD models.
The dotted lines show the combined flux from each WD plus 1 Gyr old 25, 20, 15, 10, 7, 5, 2, and 1 $M_J$
companions (from top to bottom), respectively. None of the WDs in Figure 2 show significant flux excesses
in the mid-infrared. We rule out planets more massive than 5, 7, or 10 $M_J$
around these targets with 3$\sigma$ confidence.

Table 3 presents the search radius and the unresolved companion limits for each star.
Our two pixel search radius in IRAC images corresponds to 50 AU for the nearest, and 260 AU for the most distant WD in our sample.
We expect the mass loss process to enlarge the planetary orbits by a factor of $M_{\rm MS}/M_{\rm WD}$ \citep{zuckerman87}.
We use the initial-final mass relation of \citet{kalirai08} to estimate the ratio of MS and WD masses,
i.e., the orbital expansion factor.
The above search radii allow us to probe the WDs' progenitors for planets to 11 AU for the nearest WD
and to 55 AU for the most distant one.
For the separations probed (see Table 3) at 13 of our targets,
we do not find any planets more massive than $10M_J$, according to \citet{burrows03} models.
For five of our targets,
there are no planets more massive than $5M_J$ within the probed
inner regions.

We extend our search to partially-resolved companions by increasing the size of the
photometric aperture used in our analysis.
Also, it may be possible to exclude resolved companions at larger radii by the lack of 4.5 $\mu$m point sources around
our targets. One of our targets, PG 1307+354, has two nearby sources that contaminate the photometry in apertures larger
than two pixels. For the remaining 13 targets, the differences between three pixel aperture photometry and two pixel aperture photometry
are relatively small ($\leq3\%$).
Increasing the search radius to five or ten pixel apertures is problematic as our targets are relatively
faint in the mid-infrared and there are many nearby faint sources. 

The direct detection of three planets at 24, 38, and 68 AU around HR 8799 \citep{marois08} and the detection of a planet at 119 AU
around Fomalhaut \citep{kalas08} demonstrate that giant planets exist at large separations from their host stars.
The planets around HR 8799 are massive enough ($5-13M_J$) to be detected around some of our targets.
Using a three pixel aperture and accounting for orbital expansion, we would have detected a massive planet at 24 AU
around 13 of our targets, a planet at 38 AU around
six of our targets, and a planet at 68 AU around only three of our targets.

\section{DISCUSSION}

None of the stars in our sample show mid-infrared flux excess from brown dwarfs or planetary
mass companions. Our data rule out $\geq5 M_J$ companions for
five WDs and $\geq10 M_J$ companions for 13 WDs.

There have been many other searches for substellar and planetary mass companions to WDs. In fact,
the first candidate brown dwarf was found around a WD more than 20 years ago \citep{becklin88}. However, only
a few more WD + brown dwarf systems have been discovered since then \citep{farihi04,maxted06,steele09}.
\citet{farihi05} find that 
less than 0.5\% of WDs have brown dwarf companions.

\citet{hogan09} performed a $J-$band proper motion survey of 23 WDs with Gemini, and
found that $\leq$5\% of WDs have substellar companions. 
In addition, the near- and mid-infrared searches by \citet{debes05,debes07} and \citet{friedrich06} did not reveal
any substellar companions to WDs. {\em Spitzer} IRAC currently provides the best opportunity to detect
planetary mass companions around WDs. An IRAC survey of 124 nearby WDs by \citet{mullally07}
did not find any planets. However,
accurate mass and age estimates are not available for the majority
of the stars in their sample, and a detailed analysis would be
required to put reliable limits on possibly hidden companions.

\citet{farihi08} present IRAC observations of 48 WDs including 31 WDs younger than 1 Gyr
(MS + WD cooling age). They use blackbody models to predict the 4.5 $\mu$m photospheric flux from their targets
and search for 15\% excess flux at 4.5 $\mu$m. Of course, the use of WD-atmosphere models (as in our study)
would be more accurate, but blackbody models are sufficient for finding 15\% excesses
around relatively hot WDs.
They rule out $\geq10 M_J$ companions around 27 of their targets.
The addition of 13 stars presented in this paper brings the total sample size to 40.
None of the 40 stars in the combined sample
show infrared excess due to substellar companions more massive
than $10M_J$. For a limit of $5M_J$, there are a
total of 10 WDs in both studies.
While no planets are detected (i.e., $f=0$), it remains conceivable that the frequency is non-zero.

Due to small sample size, we use a binomial probability distribution to derive statistical uncertainties.
The probability, $P(f)$, that a survey of $N$ stars will detect $n$ companions, when the true frequency of companions
is $f$ is given by \citep{burgasser03,mccarthy04}:

\begin{equation} 
P_n(f) = f^n (1-f)^{N-n} \frac{N!}{(N-n)!n!}.
\end{equation}

For $N=40$ and $n=0$, the probability distribution peaks at zero.
Since the distribution is not symmetric about its maximum value,
we report the range in frequency that delimits 34\% and 68\% of the integrated probability function as
the mean frequency and error bars, respectively.
These error bars are equivalent to 1$\sigma$ limits for a Gaussian distribution.
We find that the frequency of $\geq10 M_J$ companions to WDs is $1.0^{+1.7}_{-1.0}$\%.
This is consistent with zero.

Figure 3 displays planet versus host star mass for all known extrasolar planets detected by the radial velocity,
astrometric, transit, microlensing, and direct imaging searches as of 2009 April (based on the
Extrasolar Planets Encyclopedia\footnote{http://exoplanet.eu/}).
The parameter space that is ruled
out by IRAC observations of the WD remnants of intermediate-mass stars presented in this work and \citet{farihi08} is also shown. 
We note that this figure excludes seven stars with $M_{WD}>1.1M_\odot$ from the \citet{farihi08} study. The initial-final mass relation
of \citet{kalirai08} implies that these WDs are the descendants of stars with $M\geq6.5M_\odot$.
However, the initial-final mass relation is steep near the top end
of the WD mass function \citep{williams09}, and the errors are
relatively large. The initial-final mass relation of \citet{williams09} implies an initial-mass of $\approx6M_\odot$ for a $1.1M_\odot$
WD. Even though these seven stars are not shown in Figure 3, they are included in our statistical analysis of 40 stars presented above.

About 60\% of the Sun-like ($0.7-1.3\ M_\odot$) stars with RV detected planets have planet/host star mass ratios
$\geq 1 \frac{M_J}{ M_\odot}$ (based on the Extrasolar Planets Encyclopedia). This fraction goes down to 33\% for
planet/host star mass ratios greater than $2\frac{M_J}{ M_\odot}$.
\citet{johnson07} find that the Jovian-planet frequency is
4.2\% for Sun-like stars and it is
8.9\% for $1.3-1.9M_\odot$ stars. This trend is expected to continue for higher mass stars up to $3-4M_\odot$.
\citet{kennedy08} predict that about 20\% of $3-4\ M_\odot$ stars have at least one gas giant planet.
They also find that the isolation masses are larger for more massive stars, and
the giant planets that eventually form should also be bigger \citep[see Fig. 2 and 3 in][]{kennedy08}.

Assuming that the planet/host star mass ratios are similar for Sun-like and intermediate mass stars, we expect
6.6\% of $3-4M_\odot$ stars to have planets with planet/host star mass ratios greater than $2\frac{M_J}{ M_\odot}$.
We probe this regime for 6 stars in our sample. \citet{farihi08} reach the same limit for 20 stars, including 10 stars with
$M_{WD}>1.1M_\odot$. All 26 stars have $M_{MS}\geq3M_\odot$.
The probability of a null detection among a sample of 26 stars when the (predicted) frequency is 6.6\% is 
17\% (see equation 1). Thus, the null detections reported in this paper and \citet{farihi08} are
suggestive, but do not directly conflict with previous
detections of planets around intermediate-mass stars.
Our sample is not large enough to put more stringent limits on the frequency of planets around
intermediate-mass stars. A detailed analysis of a larger IRAC survey of WDs like that of \citet{mullally07} will provide useful
constraints on the frequency of massive planets around intermediate-mass stars.

If the current trend in the lack of discovery of planets around WDs continues, it would suggest one of the following:

1 - Intermediate- and high-mass stars may be inefficient in forming giant planets. For stars more massive than $3M_\odot$,
irradiation overcomes accretion as the stars reach the main sequence relatively quickly. This pushes the snow line to
$10-15$ AU and makes formation of cores difficult \citep{ida05,kennedy08}. 

2 - The planets may not survive the red giant and the AGB phases.
\citet{villaver07} find that planets within 5.3 AU of 5 $M_\odot$ stars
will be engulfed during the AGB phase.
Of course, avoiding engulfment may not be enough for a planet to survive because a
corpuscular drag resulting from its interaction with the stellar wind would decrease the planetary orbits. However, \citet{duncan98}
show that the planets will only move inward slightly due to the corpuscular drag.
\citet{villaver07} suggest that the maximum stellar radius is reached only for a brief period of time, and the planetary
orbits would have been expanded due to mass loss by that time. Hence, they find that the tidal drag forces are negligible
for orbital radii larger than the maximum stellar radius reached in the AGB phase.
Since the search radii for planets around the progenitors of our sample of WDs ranges from 11 to 55 AU, the engulfment of the planets
during the AGB phase cannot explain the apparent lack of high mass planets around our targets.

3 - The orbits of the planets around massive stars may become unstable during the late stages of
stellar evolution.
\citet{debes02} suggest that planets around WDs may become unstable to close
approaches with each other and the entire system may become dynamically young.
\citet{duncan98} find that the crossing times for planetary orbits depend on ($M_{\rm WD} / M_{\rm MS})^{3.2}$.
Even though this timescale is relatively long for the Sun ($\sim 10$ Gyr), the larger mass-reduction factor for more massive stars
means that the timescale for unstable close approaches is less than 1 Gyr for planets around $M \geq 3.6 M_\odot$ stars.
In addition, since orbital semi-major axes grow during the mass loss phase, some planets will pass through resonances that could create
or enhance instabilities in the system.
However, detailed simulations including the effects of orbital expansion on the stability of planets around massive WDs
are not currently available.

\section{CONCLUSIONS}

There is only one known WD that has
a possible planetary companion, GD 66. The signal from an $M \sin i = 2.4 M_J$ planetary companion
in a 4.5 yr orbit is detected in timing measurements of this pulsating WD star \citep{mullally09}.
However, a complete orbit has not been observed yet,
and the detection remains provisional.
{\em Spitzer} IRAC observations of GD 66 did not reveal any significant mid-infrared flux excess. Based on the substellar cooling models by
\citet{burrows03} and looser detection criteria compared to \citet{farihi08} and our study, \citet{mullally09} place an upper limit
of $5-7 M_J$ on the mass of the companion. 

So far, no other search, including the IRAC surveys by \citet{mullally07}, \citet{farihi08}, and this paper has been succesful
in finding planetary companions to WDs. The proper motion surveys in
the near-infrared have ruled out resolved companions to two dozen WDs \citep{hogan09}. However, a proper
motion survey of the majority of the WDs observed with IRAC has not been performed yet. Such a survey will be valuable
for searching for resolved companions to WDs. Future studies with ground based telescopes using Adaptive Optics instruments and
with the {\em James Webb Space Telescope (JWST)} may provide the first answers to whether Jupiter mass planets survive around WDs or not.

\acknowledgements
Support for this work was provided by NASA through the Spitzer Space Telescope Fellowship Program,
under an award from Caltech. A.G. was supported by NSF grant AST-0757888.
We thank the referee, J. Farihi, for a detailed and constructive report. We also thank S. Kenyon for a careful
reading of this manuscript.

\clearpage
\begin{deluxetable}{lrrrrrrr}
\tabletypesize{\footnotesize}
\tablecolumns{8}
\tablewidth{0pt}
\tablecaption{High Mass WD Targets}
\tablehead{
\colhead{Object}&
\colhead{$T_{\rm eff}$}&
\colhead{$\log g$}&
\colhead{$M_{\rm WD}$}&
\colhead{$M_{\rm MS}$}&
\colhead{$d$}&
\colhead{$\tau_{\rm WD}$}&
\colhead{$\tau_{\rm WD+MS}$}
\\
        & (K) & (cm s$^{-2}$) & ($M_\odot$) & ($M_\odot$) & (pc) & (Myr) & (Myr)}
\startdata
PG 0852+659 & 19070 & 8.13 & 0.70 & 2.8 & 92 & 140 & 900 \\
PG 1034+492 & 20650 & 8.17 & 0.73 & 3.1 & 80 & 120 & 720 \\
PG 1038+634 & 24450 & 8.38 & 0.87 & 4.4 & 68 & 100 & 350 \\
PG 1051+274 & 23100 & 8.37 & 0.86 & 4.3 & 41 & 110 & 380 \\
PG 1108+476 & 12400 & 8.31 & 0.80 & 3.7 & 46 & 600 & 980 \\
PG 1129+156 & 16890 & 8.19 & 0.73 & 3.1 & 36 & 220 & 830 \\
PG 1201$-$001& 19770 & 8.26 & 0.78 & 3.5 & 63 & 160 & 580 \\
PG 1307+354 & 11180 & 8.15 & 0.70 & 2.8 & 45 & 630 & 1390 \\
PG 1310+583 & 10560 & 8.32 & 0.80 & 3.7 & 21 & 910 & 1290 \\
PG 1319+466 & 13880 & 8.19 & 0.73 & 3.1 & 37 & 400 & 1000 \\
PG 1335+701 & 30140 & 8.25 & 0.79 & 3.6 & 108 & 30 & 420 \\
PG 1446+286 & 22890 & 8.42 & 0.89 & 4.6 & 47 & 130 & 350 \\
PG 1515+669 & 10320 & 8.40 & 0.86 & 4.3 & 33 & 1120 & 1390 \\
PG 1550+183 & 14260 & 8.25 & 0.77 & 3.5 & 41 & 390 & 840 
\enddata
\end{deluxetable}

\begin{deluxetable}{lccccc}
\tabletypesize{\footnotesize}
\tablecolumns{5}
\tablewidth{0pt}
\tablecaption{IRAC Photometry of High Mass WDs}
\tablehead{
\colhead{Object}&
\colhead{3.6$\mu$m}&
\colhead{4.5$\mu$m}&
\colhead{5.8$\mu$m}&
\colhead{8.0$\mu$m}&
\colhead{Reduction}
\\
        & ($\mu$Jy) & ($\mu$Jy) & ($\mu$Jy) & ($\mu$Jy) & Pipeline}
\startdata
PG 0852+659 & 97.7 $\pm$ 3.9 & 65.1 $\pm$ 3.3 & 44.3 $\pm$ 13.1 & 25.2 $\pm$ 10.6 & 17.2\\
PG 1034+492 & 93.5 $\pm$ 5.6 & 57.5 $\pm$ 3.1 & 40.9 $\pm$ 5.9 & 31.7 $\pm$ 20.3 & 18.5\\
PG 1038+634 & 107.3 $\pm$ 4.0 & 65.8 $\pm$ 2.7 & 50.7 $\pm$ 17.4 & 31.0 $\pm$ 16.6 & 18.5\\
PG 1051+274 & 299.1 $\pm$ 12.1 & 190.6 $\pm$ 8.7 & 126.0 $\pm$ 25.5 & 59.1 $\pm$ 31.2 & 18.5\\
PG 1108+476 & 138.4 $\pm$ 5.5 & 87.3 $\pm$ 4.1 & 69.9 $\pm$ 15.9 & 34.6 $\pm$ 26.4 & 17.2\\
PG 1129+156 & 371.8 $\pm$ 12.3 & 230.8 $\pm$ 11.4 & 154.6 $\pm$ 22.1 & 74.9 $\pm$ 19.1 & 18.5\\
PG 1201$-$001 & 123.1 $\pm$ 4.7 & 79.1 $\pm$ 3.9 & 50.9 $\pm$ 9.0 & 28.9 $\pm$ 24.8 & 18.5\\
PG 1307+354 & 191.8 $\pm$ 6.3 & 118.4 $\pm$ 6.7 & 95.7 $\pm$ 23.6 & 60.5 $\pm$ 25.8 & 18.5\\
PG 1310+583 & 658.5 $\pm$ 23.0 & 420.2 $\pm$ 19.6 & 287.4 $\pm$ 22.4 & 159.3 $\pm$ 29.6 & 17.2\\
PG 1319+466 & 279.9 $\pm$ 11.2 & 177.3 $\pm$ 6.9 & 110.8 $\pm$ 25.2 & 77.9 $\pm$ 24.3 & 18.5\\
PG 1335+701 & 92.1 $\pm$ 3.6 & 56.5 $\pm$ 3.3 & 30.4 $\pm$ 9.6 & 24.5 $\pm$ 11.0 & 17.2\\
PG 1446+286 & 195.1 $\pm$ 9.7 & 120.1 $\pm$ 6.2 & 99.7 $\pm$ 18.6 & 54.5 $\pm$ 24.7 & 18.5\\
PG 1515+669 & 187.7 $\pm$ 7.9 & 119.4 $\pm$ 7.5 & 94.4 $\pm$ 16.9 & 24.4 $\pm$ 11.9 & 17.2\\
PG 1550+183 & 218.3 $\pm$ 8.0 & 136.6 $\pm$ 5.3 & 96.0 $\pm$ 20.7 & 50.3 $\pm$ 11.7 & 18.5  
\enddata
\end{deluxetable}

\clearpage
\begin{deluxetable}{lrccc}
\tabletypesize{\footnotesize}
\tablecolumns{5}
\tablewidth{0pt}
\tablecaption{Limits on Unresolved Companions}
\tablehead{
\colhead{Object}&
\colhead{Search Radius\tablenotemark{\dagger}}&
\colhead{Initial Separation\tablenotemark{\dagger}}&
\colhead{$M_{\rm MS}$}&
\colhead{Ruled out\tablenotemark{\dagger\dagger}}
\\   & (AU) & (AU) & ($M_\odot$) & Companions}
\startdata
PG 0852+659 & 221 & 55 & 2.8 & $> 15 M_J$ \\
PG 1034+492 & 192 & 46 & 3.1 & $> 7  M_J$ \\
PG 1038+634 & 163 & 32 & 4.4 & $\geq 5 M_J$ \\
PG 1051+274 & 98  & 20 & 4.3 & $\geq 5 M_J$ \\
PG 1108+476 & 110 & 24 & 3.7 & $\geq 5 M_J$ \\
PG 1129+156 & 86  & 20 & 3.1 & $> 7  M_J$ \\
PG 1201$-$001& 151 & 33 & 3.5 & $\geq 10 M_J$ \\
PG 1307+354 & 108 & 27 & 2.8 & $> 10  M_J$ \\	
PG 1310+583 & 50  & 11 & 3.7 & $\geq 10 M_J$ \\
PG 1319+466 & 89  & 21 & 3.1 & $\geq 5 M_J$ \\
PG 1335+701 & 259 & 56 & 3.6 & $\geq 10 M_J$ \\
PG 1446+286 & 113 & 22 & 4.6 & $\geq 5 M_J$ \\
PG 1515+669 & 79  & 16 & 4.3 & $\geq 7 M_J$ \\
PG 1550+183 & 98  & 22 & 3.5 & $\geq 7 M_J$ \\
\enddata
\tablenotetext{\dagger}{For a two pixel aperture.}
\tablenotetext{\dagger\dagger}{These limits
are calculated based on the models by \citet{burrows03}. See the discussion in Section 3.}
\end{deluxetable}

\clearpage
\begin{figure}
\hspace{-0.8in}
\includegraphics[angle=-90,scale=.75]{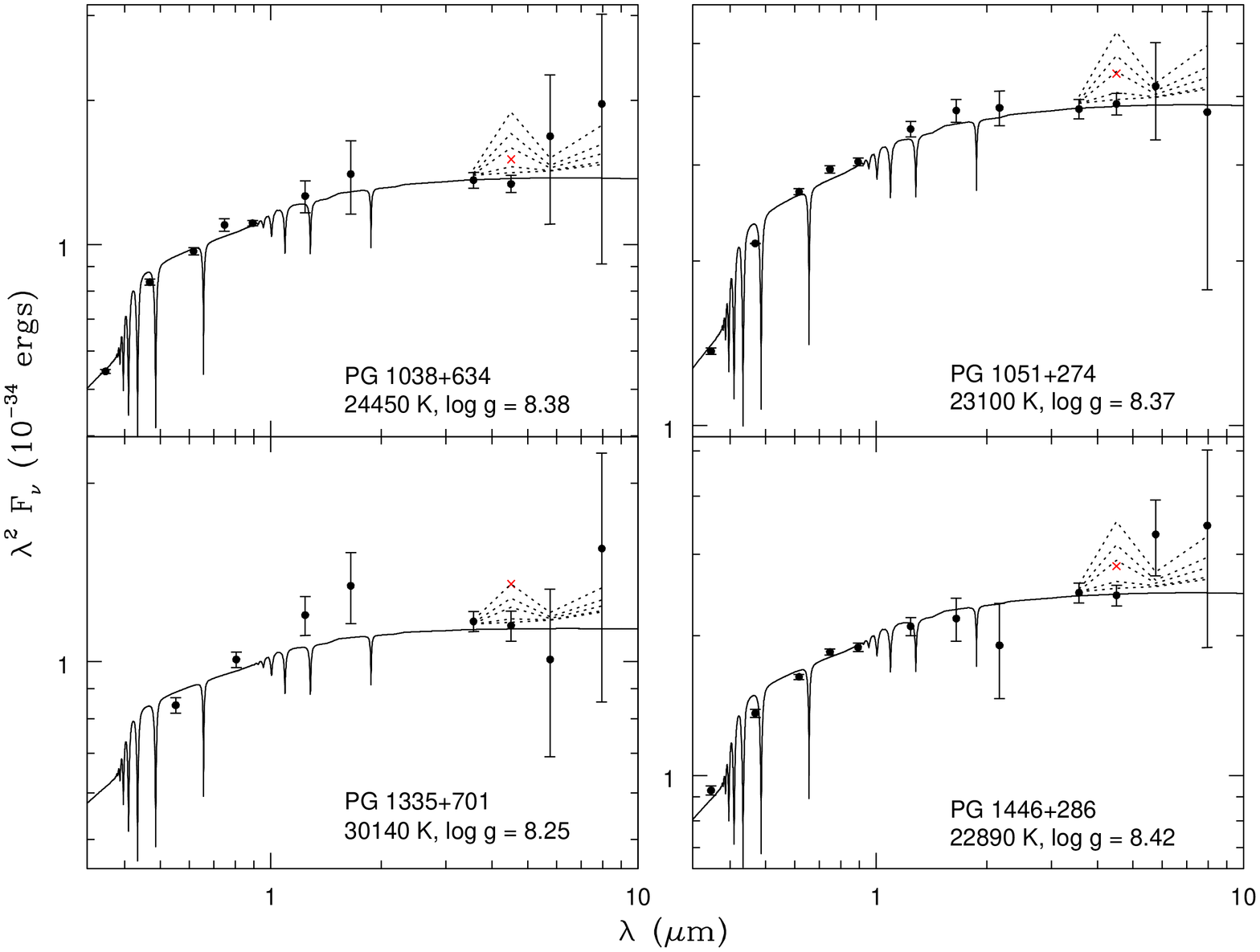}
\caption{Spectral energy distributions of four 300 Myr old (total age) WDs. The expected photospheric flux
from the WDs are shown as solid lines \citep{koester09}. The dotted lines show the expected flux
from planetary companions with $M=10, 7, 5, 2$ and 1 $M_J$ (from top to bottom), respectively.
The red cross marks the 3$\sigma$ upper limit of the 4.5$\mu$m photometry.}
\end{figure}

\clearpage
\begin{figure}
\hspace{-0.8in}
\includegraphics[angle=-90,scale=.75]{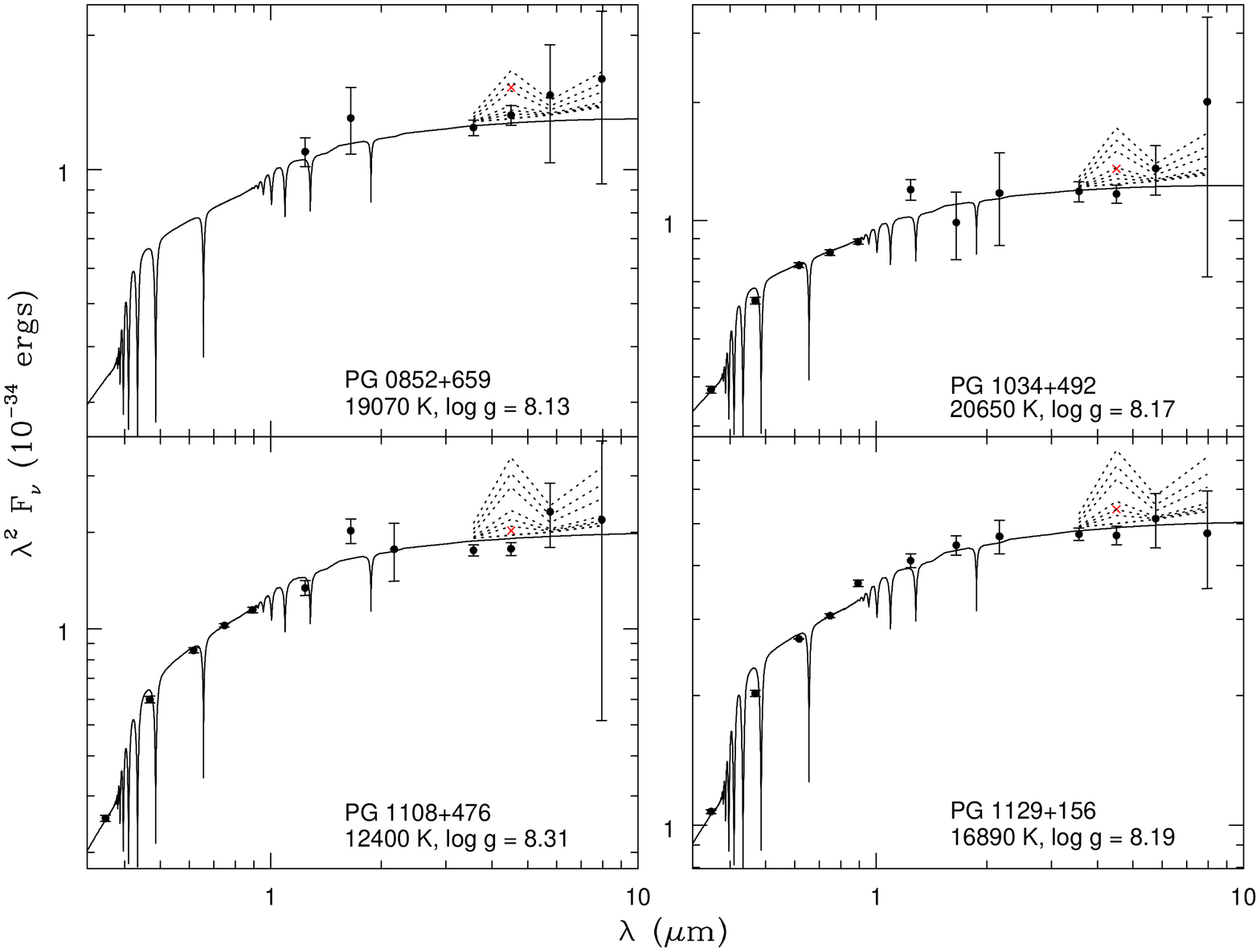}
\caption{Spectral energy distributions of $\approx$1 Gyr old (total age) WDs. The expected photospheric flux
from the WDs are shown as solid lines \citep{koester09}. The dotted lines show the expected flux
from planetary companions with $M=25, 20, 15, 10, 7, 5, 2$ and 1 $M_J$ (from top to bottom), respectively.
The red cross marks the 3$\sigma$ upper limit of the 4.5$\mu$m photometry.}
\end{figure}

\clearpage
\begin{figure}
\hspace{-0.8in}
\includegraphics[angle=-90,scale=.75]{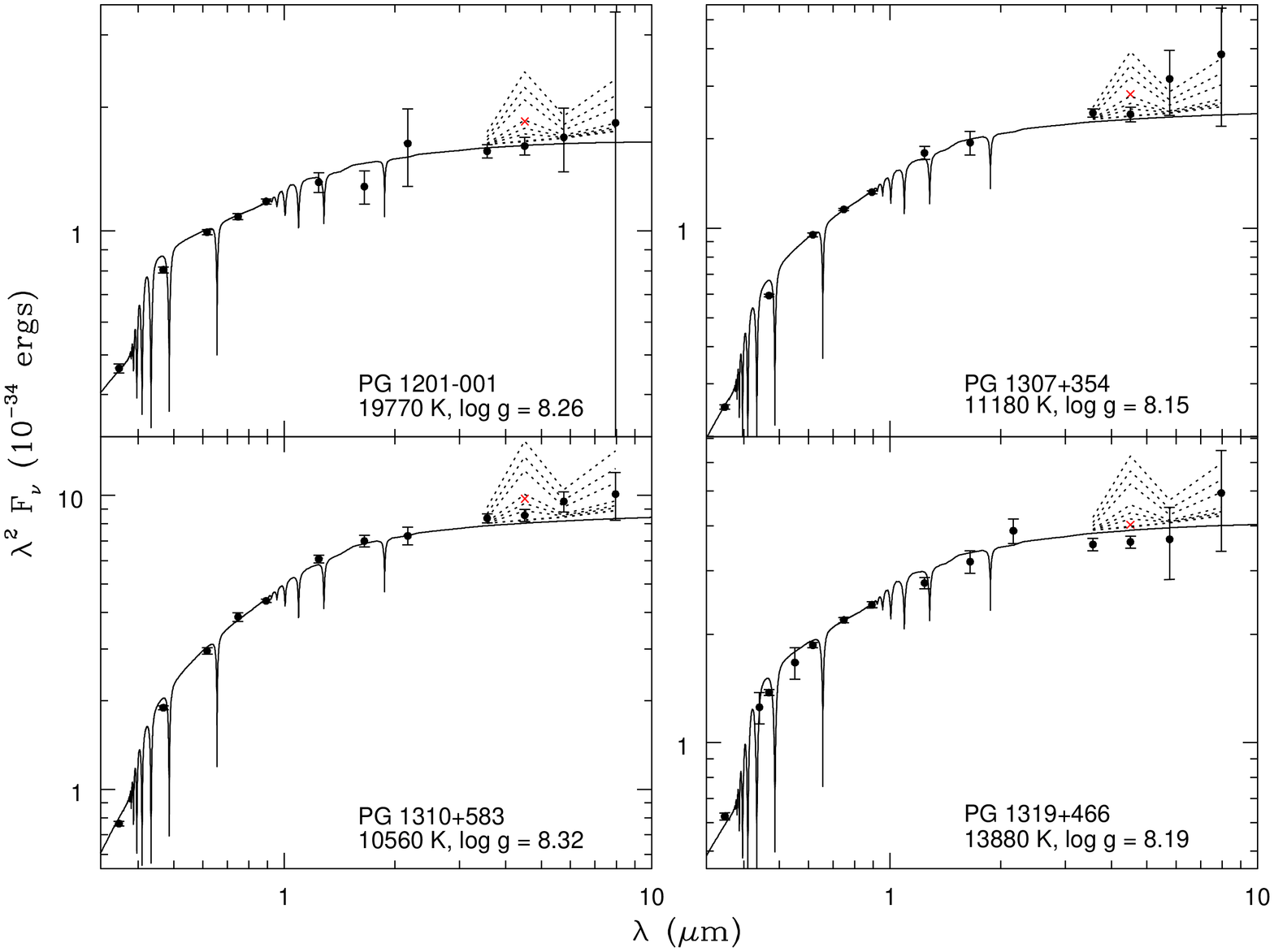}
\figurenum{2}
\caption{contd.}
\end{figure}

\clearpage
\begin{figure}
\plotone{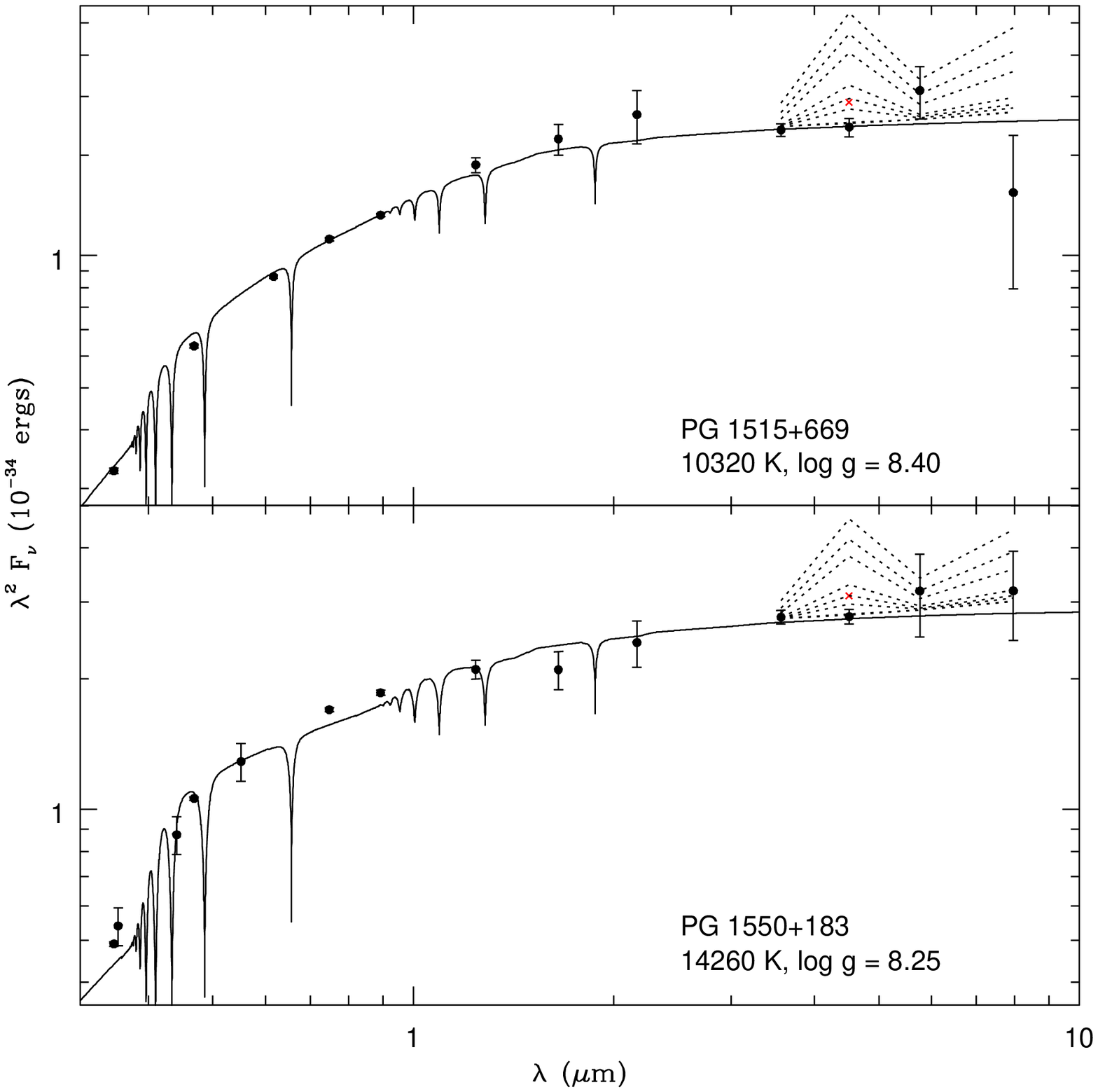}
\figurenum{2}
\caption{contd.}
\end{figure}

\clearpage
\begin{figure}
\plotone{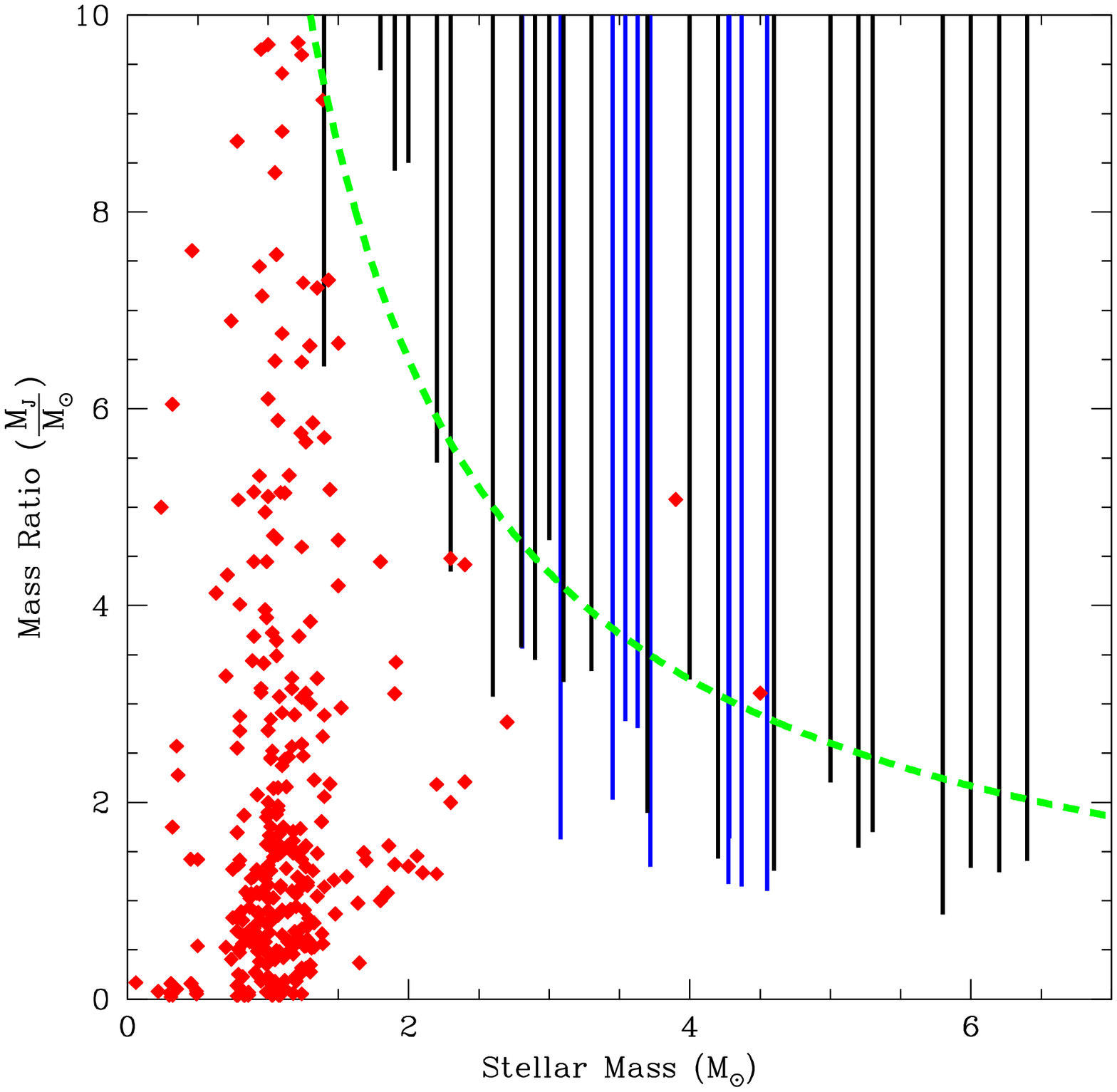}
\caption{Planet versus host mass for all known extrasolar planets detected by
the RV, astrometry, transits, microlensing, and direct imaging observations as
of 2009 April. The dashed line marks the upper limit for planetary mass objects ($M=13 M_J$).
The parameter space that is ruled out by {\em Spitzer} observations
of WD remnants of intermediate-mass stars by \citet{farihi08} and this work is shown as black and blue lines, respectively.
We use 3$\sigma$ (12-18\%), while \citet{farihi08} use 15\% excess for planet mass upper limits. 
}
\end{figure}


\begin{thebibliography}{}
\bibitem[Becklin \& Zuckerman(1988)]{becklin88} Becklin, E.~E., \& Zuckerman, B.\ 1988, \nat, 336, 656
\bibitem[Burgasser et al.(2003)]{burgasser03} Burgasser, A.~J., Kirkpatrick, J.~D., Reid, I.~N., Brown, M.~E., Miskey, C.~L., \& Gizis, J.~E.\ 2003, \apj, 586, 512
\bibitem[Burleigh et al.(2002)]{burleigh02} Burleigh, M.~R., Clarke, F.~J., \& Hodgkin, S.~T.\ 2002, \mnras, 331, L41
\bibitem[Burrows et al.(2003)]{burrows03} Burrows, A., Sudarsky, D., \& Lunine, J.~I.\ 2003, \apj, 596, 587
\bibitem[Burrows et al.(2008)]{burrows08} Burrows, A., Budaj, J., \& Hubeny, I. 2008, \apj, 678, 1436
\bibitem[Debes \& Sigurdsson (2002)]{debes02} Debes, J. H. \& Sigurdsson, S. 2002, \apj, 572, 556
\bibitem[Debes et al.(2005)]{debes05} Debes, J.~H., Sigurdsson, S., \& Woodgate, B.~E.\ 2005, \apj, 633, 1168
\bibitem[Debes et al.(2007)]{debes07} Debes, J.~H., Sigurdsson, S., \& Hansen, B.\ 2007, \aj, 134, 1662
\bibitem[Duncan \& Lissauer(1998)]{duncan98} Duncan, M.~J., \& Lissauer, J.~J.\ 1998, Icarus, 134, 303
\bibitem[Eisenstein et al.(2006)]{eisenstein06} Eisenstein, D.~J., et al.\ 2006, \apjs, 167, 40
\bibitem[Farihi \& Christopher(2004)]{farihi04} Farihi, J., \& Christopher, M.\ 2004, \aj, 128, 1868
\bibitem[Farihi et al.(2005)]{farihi05} Farihi, J., Becklin, E.~E., \& Zuckerman, B.\ 2005, \apjs, 161, 394
\bibitem[Farihi et al.(2008)]{farihi08} Farihi, J., Becklin, E.~E., \& Zuckerman, B.\ 2008, \apj, 681, 1470
\bibitem[Farihi et al.(2009)]{farihi09} Farihi, J., Jura, M., \& Zuckerman, B.\ 2009, \apj, 694, 805
\bibitem[Fischer \& Valenti(2005)]{fischer05} Fischer, D.A. \& Valenti, J. 2005, \apj, 622, 1102
\bibitem[Friedrich et al.(2006)]{friedrich06} Friedrich, S., Zinnecker, H., Correia, S., Brandner, W., Burleigh, M., \& McCaughrean, M. 2006, ASP Conference Series, 999
\bibitem[Gould \& Kilic(2008)]{gould08} Gould, A., \& Kilic, M.\ 2008, \apjl, 673, L75
\bibitem[Gould(2009)]{gould09} Gould, A.\ 2009, Astronomical Society of the Pacific Conference Series, 403, 86
\bibitem[Hogan et al.(2009)]{hogan09} Hogan, E., Burleigh, M. R., \& Clarke, F. J. 2009, \mnras, in press
\bibitem[Hubeny \& Burrows(2007)]{hubeny07} Hubeny, I., \& Burrows, A.\ 2007, \apj, 669, 1248
\bibitem[Ida \& Lin(2005)]{ida05} Ida, S., \& Lin, D.~N.~C.\ 2005, \apj, 626, 1045
\bibitem[Ignace(2001)]{ignace01} Ignace, R.\ 2001, \pasp, 113, 1227
\bibitem[Johnson et al.(2007)]{johnson07} Johnson, J.A., Butler, R.P., Marcy, G.W., Fischer, D.A., Vogt, S.S., Wright, J.T., \& Peek, K.M.G., 2007, \apj, 670, 833
\bibitem[Jura(2003)]{jura03} Jura, M.\ 2003, \apjl, 584, L91
\bibitem[Jura(2006)]{jura06} Jura, M.\ 2006, \apj, 653, 613 
\bibitem[Jura et al.(2009)]{jura09} Jura, M., Farihi, J., \& Zuckerman, B.\ 2009, \aj, 137, 3191
\bibitem[Kalas et al.(2008)]{kalas08} Kalas, P., et al.\ 2008, Science, 322, 1345
\bibitem[Kalirai et al.(2008)]{kalirai08} Kalirai, J.~S., Hansen, B.~M.~S., Kelson, D.~D., Reitzel, D.~B., Rich, R.~M., \& Richer, H.~B.\ 2008, \apj, 676, 594
\bibitem[Kennedy \& Kenyon(2008)]{kennedy08} Kennedy, G.~M., \& Kenyon, S.~J.\ 2008, \apj, 673, 502
\bibitem[Kilic et al.(2008)]{kilic08} Kilic, M., Farihi, J., Nitta, A., \& Leggett, S.~K.\ 2008, \aj, 136, 111
\bibitem[Koester (2009)]{koester09} Koester, D. 2009, Mem. Soc. Astron. Italiana, in press
\bibitem[Koester et al.(2005)]{koester05} Koester, D., Rollenhagen, K., Napiwotzki, R., Voss, B., Christlieb, N., Homeier, D., \& Reimers, D.\ 2005, \aap, 432, 1025
\bibitem[Landolt(1992)]{landolt92} Landolt, A.~U.\ 1992, \aj, 104, 340
\bibitem[Laws et al.(2003)]{laws03}Laws, C., Gonzalez, G., Walker, K.M., Tyagi, S., Dodsworth, J., Snider, K., \& Suntzeff, N.B. 2003, \aj, 125, 2664
\bibitem[L{\'e}pine \& Shara(2005)]{lepine05} L{\'e}pine, S., \& Shara, M.~M.\ 2005, \aj, 129, 1483
\bibitem[Liebert et al.(2005)]{liebert05} Liebert, J., Bergeron, P., \& Holberg, J.~B.\ 2005, \apjs, 156, 47
\bibitem[Livio et al.(2005)]{livio05} Livio, M., Pringle, J.~E., \& Wood, K.\ 2005, \apjl, 632, L37
\bibitem[Marcy et al.(2005)]{marcy05} Marcy, G., Butler, R.~P., Fischer, D., Vogt, S., Wright, J.~T., Tinney, C.~G., \& Jones, H.~R.~A.\ 2005, Progress of Theoretical Physics Supplement, 158, 24
\bibitem[Marois et al.(2008)]{marois08} Marois, C., Macintosh, B., Barman, T., Zuckerman, B., Song, I., Patience, J., Lafreni{\`e}re, D., \& Doyon, R.\ 2008, Science, 322, 1348
\bibitem[Maxted et al.(2006)]{maxted06} Maxted, P.~F.~L., Napiwotzki, R., Dobbie, P.~D., \& Burleigh, M.~R.\ 2006, \nat, 442, 543
\bibitem[Mazeh(2009)]{mazeh09} Mazeh, T.\ 2009, IAU Symposium, 253, 11
\bibitem[McCarthy \& Zuckerman(2004)]{mccarthy04} McCarthy, C., \& Zuckerman, B.\ 2004, \aj, 127, 2871
\bibitem[Mullally et al.(2007)]{mullally07} Mullally, F., Kilic, M., Reach, W.~T., Kuchner, M.~J., von Hippel, T., Burrows, A., \& Winget, D.~E.\ 2007, \apjs, 171, 206
\bibitem[Mullally et al.(2009)]{mullally09} Mullally, F., Reach, W.~T., Degennaro, S., \& Burrows, A.\ 2009, \apj, 694, 327
\bibitem[Reach et al.(2005)]{reach05} Reach, W.~T., et al.\ 2005, \pasp, 117, 978
\bibitem[Silvotti et al.(2007)]{silvotti07} Silvotti, R., et al.\ 2007, \nat, 449, 189
\bibitem[Steele et al.(2009)]{steele09} Steele, P.~R., Burleigh, M.~R., Farihi, J., G{\"a}nsicke, B.~T., Jameson, R.~F., Dobbie, P.~D., \& Barstow, M.~A.\ 2009, \aap, 500, 1207
\bibitem[Udry \& Santos(2007)]{udry07} Udry, S., \& Santos, N.~C.\ 2007, \araa, 45, 397
\bibitem[Villaver \& Livio(2007)]{villaver07} Villaver, E., \& Livio, M.\ 2007, \apj, 661, 1192
\bibitem[Williams et al.(2009)]{williams09} Williams, K.~A., Bolte, M., \& Koester, D.\ 2009, \apj, 693, 355
\bibitem[Wood (1992)]{wood92} Wood, M. A. 1992, \apj, 386, 539
\bibitem[Zuckerman \& Becklin(1987)]{zuckerman87} Zuckerman, B., \& Becklin, E.~E.\ 1987, \apjl, 319, L99
\bibitem[Zuckerman et al.(2003)]{zuckerman03} Zuckerman, B., Koester, D., Reid, I.~N., \& H{\"u}nsch, M.\ 2003, \apj, 596, 477
\bibitem[Zuckerman et al.(2007)]{zuckerman07} Zuckerman, B., Koester, D., Melis, C., Hansen, B.~M., \& Jura, M.\ 2007, \apj, 671, 872
\end{thebibliography}
\end{document}